\documentclass[aps,amsmath,reprint,amssymb,prb]{revtex4-1}
\usepackage{graphicx,color,bm,amsmath,url,natbib}

\usepackage[utf8]{inputenc}
\usepackage{mathtools}

\begin{document}

\title{Rotational transition, domain formation, dislocations and defects in vortex systems with combined six- and 12-fold anisotropic interactions}

\author{M.~W.~Olszewski}
\affiliation{Department of Physics, University of Notre Dame, Notre Dame, Indiana 46656, USA}

\author{M.~R.~Eskildsen}
\altaffiliation{Corresponding author: eskildsen@nd.edu}
\affiliation{Department of Physics, University of Notre Dame, Notre Dame, Indiana 46656, USA}

\author{C.~Reichhardt}
\author{C.~J.~O.~Reichhardt}
\affiliation{Theoretical Division, Los Alamos National Laboratory, Los Alamos, New Mexico 87545, USA}

\date{\today}

\begin{abstract}
We introduce a phenomenological model for a pairwise repulsive interaction potential of vortices in a type-II superconductor,  consisting of superimposed six- and 12-fold anisotropies.
Using numerical simulations we study how the vortex lattice configuration varies as the magnitudes of the two anisotropic interaction terms change.
A triangular lattice appears for all values, and rotates through 30$^\circ$ as the ratio of the six- and 12-fold anisotropy amplitudes is varied, in agreement with experimental results.
The transition causes the vortex lattice to split into domains that have rotated clockwise or counter-clockwise, with grain boundaries
that are ``decorated'' by dislocations consisting of five- and seven-fold coordinated vortices.
We also find intra-domain dislocations and defects, and characterize them in terms of their energy cost.
We discuss how this model could be generalized to other particle-based systems with anisotropic interactions,
such as colloids, and consider the limit of very large anisotropy where it is possible to create cluster crystal states.
\end{abstract}

\maketitle

\section{Introduction}
Magnetic flux enters a type-II superconductor in the form of quantized vortices\cite{Abrikosov:1957vu}.
The interaction between vortices is repulsive and causes them to crystallize in an ordered vortex lattice (VL), unless pinning forces or disordering due to thermal fluctuations are dominant\cite{Blatter:1994gz}.
In an ideal isotropic superconductor the VL is triangular\cite{Kleiner:1964ih} and is oriented randomly with respect to the crystal directions of the host material.
There are, however, a large class of superconductors that possess a hierarchy of anisotropies that strongly influence the VL symmetry and/or orientation relative to the crystalline axes, and which can often give rise to magnetic field or temperature driven transitions\cite{Laver:2006aa, Muhlbauer:2019jt,Laver:2009bk}.
Examples include the transition from a square to a hexagonal vortex lattice in rare earth nickelborocarbides\cite{Yaron96,Eskildsen01,Dewhurst05}, heavy fermion systems\cite{Eskildsen:2003aa,Bianchi:2008aa,White10}, and high temperature superconductors\cite{Brown04,Cameron14}.
Triangular to square vortex lattice transitions can also appear in superfluid systems\cite{Schweikhard04,Chernodub12}. 

Theoretical approaches used to study structural transitions of the VL include modifications to the London model\cite{Kogan:1997vm,Franz:1997jn,Champel:2001kf}, the addition of four-fold symmetric terms to the Ginzburg-Landau free energy\cite{Park:1998vn}, Eilenberger theory \cite{Nakai:2002aa}, modified Ginzburg-Landau approaches \cite{Klironomos:2003ks},
and modifications to the vortex interactions produced by strain fields \cite{Lin:2017tn}.
Vortex lattice ordering can also be studied using molecular dynamics (MD) methods, which treat the vortices as point particles with bulk Bessel function interactions or thin film Pearl interactions.
MD methods have been used previously to study structural vortex transitions for varied anisotropy\cite{Shi91,Olson98a,Fangohr01,Reichhardt01c,Klongcheongsan10}, but this work was limited to systems with isotropic vortex interactions due to the complexity introduced by including a fully anisotropic interaction potential, which produces nonradial forces between the vortices.

Olszewski {\it et al.} recently implemented  a phenomenological model  that makes it possible to perform MD simulations of the  VL in
the presence of four-fold anisotropic interactions, and used this model to study a square to hexagonal vortex transition\cite{Olszewski:2018fp}.
Motivated by the continuous VL rotation transitions observed in the hexagonal superconductors MgB$_2$\cite{Cubitt:2003aa,Das:2012cf} and UPt$_3$\cite{Huxley:2000aa,Avers:wx} by small-angle neutron scattering (SANS), in this work we expand the anisotropic vortex model to include a combined six- and 12-fold anisotropy.
Through MD simulations we are able to reproduce the rotation transition by varying the ratio of the six- and 12-fold contributions to the anisotropic interaction potential, validating our phenomenological model.
The competition between the six- and 12-fold anisotropies results in the formation of a rich variety of defects and domains that are distinct from what is observed in systems where the lattice defects are produced by quenched disorder.
We show how the defect structures and the range of the defect-defect interactions change across the transition from six- to 12-fold symmetry.

Another behavior that has been attracting attention is the formation of a vortex cluster lattice, which can appear when competing interactions or multiple interaction length scales are present even when the interactions are isotropic\cite{Reichhardt10,Drocco13,Komendova13,Garaud15,Meng17,Fomin19}.   
Here we demonstrate that in the limit of large anisotropy, a novel vortex cluster crystal emerges, suggesting that strong anisotropy may be a route for creating vortex clusters as well as cluster phases in particle-based systems with competing anisotropies.

\begin{figure*}
  \includegraphics{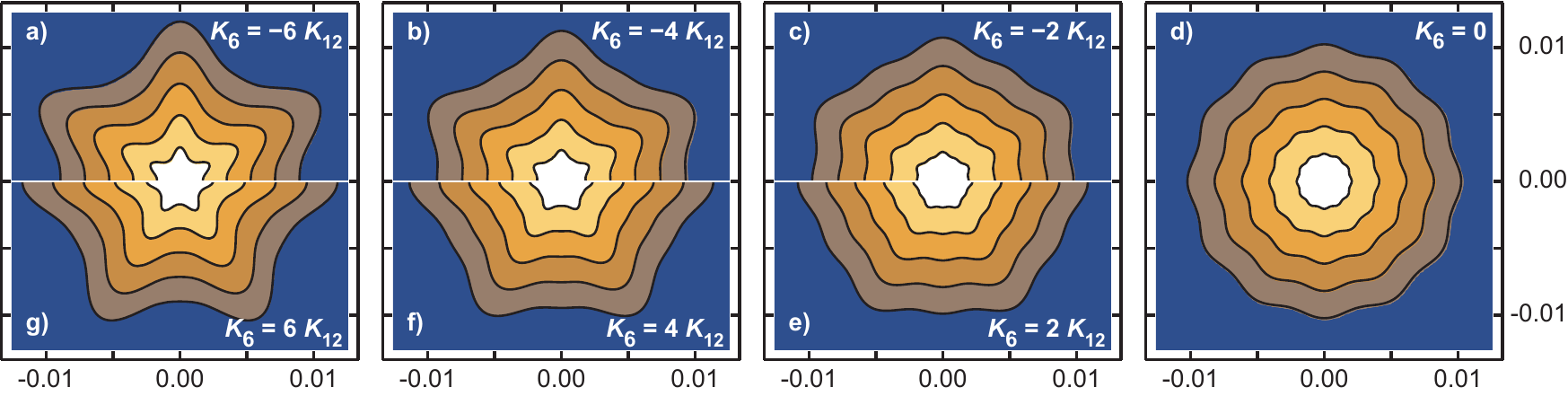}
    \caption{\label{Potentials}
      Equipotential lines for the vortex-vortex interaction in Eq.~(\ref{K612potential}) at different values of the anisotropy ratio: $\kappa = $ (a) -6, (b) -4, (c) -2, (d) 0, (e) 2, (f) 4, and (g) 6.}
\end{figure*}

In addition to the equilibrium VL configurations, MD simulations give access to the statics and dynamics of a large number of vortices over long times.
Our model can be applied to other particle-based systems with anisotropic interactions, including skyrmion lattices\cite{Jonietz:2010jz,White:2014ji,Zhang:2018bg} or colloidal particles with anisotropic interactions\cite{Eisenmann:2004iv,Glotzer:2007uo}.
It will also allow modeling of the dynamics associated with the generation and recombination of dislocations, grain boundaries and domain formation in the VL\cite{Louden:2019bq,Louden:2019wx,Louden:2019jn} as well as in general\cite{Irvine:2013dk,Lavergne:2018ds}.

\section{Methods}
\label{Methods}
The conventional model used for MD simulations of vortices in an isotropic type-II superconductor is a pairwise isotropic repulsive potential that is proportional to the zeroth order Bessel function, $U(r) = K_0(r)$.\cite{Tinkham:1996un}
Previously, we extended this model in order to make it applicable to materials with anisotropic vortex interactions \cite{Olszewski:2018fp}.
The result was the following pairwise repulsive potential:
\begin{equation}
  U(r,\theta) = A_v K_0(r)\left[ 1 + K_a \cos^2 \left( \frac{n_a (\theta-\phi_a)}{2} \right)\right]
\end{equation}
where $r = |\bm{r}_i - \bm{r}_j|$ is the distance between two vortices at positions $\bm{r}_i$ and $\bm{r}_j$.
The angle between the vortices is given with respect to the positive $x$-axis as $\theta = \tan^{-1}(r_x / r_y)$ with $\bm{r} = \bm{r}_i - \bm{r}_j$, $r_x = \bm{r} \cdot \bm{\hat{x}}$ and $r_y = \bm{r} \cdot \bm{\hat{y}}$.
The isotropic strength of the pairwise repulsion is determined by $A_v$, and is used as a normalization parameter.
The magnitude and order of the anisotropic contribution to the vortex interaction is given by $K_a$ and $n_a$ respectively, and a reference direction is specified by the angle $\phi_a$.
We used this model to study vortices in superconductors with a four-fold anisotropy ($n_a = 4$), and were able to reproduce the well-known triangular to square VL transition via an intermediate rhombic phase\cite{Olszewski:2018fp}.
In addition, we discovered the presence of ``chain states'' for large values of $K_4$ produced when deep minima in the interaction potential led to a net attractive interaction between the vortices.
This model in Eq.~(1) is applicable when only a single order of anisotropy is present.

To allow for more complex vortex interactions, we propose the following modification which includes two anisotropies of different orders, superimposed in the same potential:
\begin{widetext}
  \begin{equation}
    U(r,\theta) = A_v K_0(r) \left[ 1 + K_{\alpha} \cos^2 \left(\frac{n_{\alpha}\theta}{2} \right) + K_{\beta} \cos^2 \left( \frac{n_{\beta}\theta}{2}\right)\right]
    \label{K612potential}
  \end{equation}
\end{widetext}
where $r$, $\theta$ and $A_v$ are the same as before.
The parameters $n_{\alpha}$ and $n_{\beta}$ represent the anisotropic orders for the two anisotropies, with magnitudes given by $K_{\alpha}$ and $K_{\beta}$, respectively.
We focus on the particular case of $n_{\alpha} = 6$ and $n_{\beta} = 12$.
The interaction potential in Eq.~(\ref{K612potential}) is motivated by theoretical work, based on Ginzburg-Landau (GL) theory, describing the field and temperature dependence of the VL in the multigap superconductor MgB$_2$\cite{Zhitomirsky:2004aa}.
Here, the 12-fold term arises from an expansion of fourth-order terms in the GL functional.
Subsequent numerical studies, based on Eilenberger theory and first-principle band structure calculations for MgB$_2$\cite{Hirano:2013jx}, found that competing six-fold interactions on the different Fermi surface sheets are responsible for the continuous rotation of the triangular VL observed experimentally in this material\cite{Cubitt:2003aa,Das:2012cf}.
We note that the interaction potential in Eq.~(\ref{K612potential}) is not based on an {\em ab-initio} calculation, but rather it is a phenomenological model.
However, as shown in Sect.~\ref{rottrans}, it reproduces the experimentally observed macroscopic VL behavior for a suitable choice of parameters $K_6$ and $K_{12}$.
It can be considered as a minimal model, and it is thus reasonable also to use this expression to describe the VL in UPt$_3$ where a similar rotation transition is observed\cite{Huxley:2000aa,Avers:wx}.

For the combined six- and 12-fold anisotropy, the number and angular position of the minima and maxima in the interaction potential depends on the anisotropy ratio $\kappa = K_6/K_{12}$, as illustrated in Fig.~\ref{Potentials}.
When $|\kappa| > 4$, the potential is dominated by the six-fold term, resulting in minima along either the horizontal ($\kappa = -6$) or vertical ($\kappa = 6$) axis as well as at every $60^{\circ}$ increment from this axis, as shown in Figs.~\ref{Potentials}(a) and \ref{Potentials}(g), respectively.
Reducing the magnitude of the anisotropy ratio causes the minima to widen and become shallower, and at $\kappa = \pm 4$ the width of the minima is maximized, as indicated in Figs.~\ref{Potentials}(b) and \ref{Potentials}(f).
When $|\kappa|$ is further reduced, each local minimum splits into two minima that rotate continuously away from the high symmetry directions in the manner illustrated in Fig.~\ref{Potentials}(c) and \ref{Potentials}(e).
For $\kappa = 0$ the potential has perfect 12-fold anisotropy, with minima along the vertical axis and at every $30^{\circ}$ increment from this axis, as shown in Fig.~\ref{Potentials}(d).
For $|\kappa| \leq 4$, the angular locations of the minima in the interaction potential are given by
\begin{equation}
  \theta_{\text{min}}(\kappa) = \pm \frac{1}{6} \arccos \left( - \frac{\kappa}{4} \right).
  \label{thetamin}
\end{equation}
Here, six of the minima rotate clockwise as $\kappa$ is reduced while the other six rotate counterclockwise.
Changing the sign of $\kappa$ is equivalent to a $30^{\circ}$ rotation of the interaction potential which exchanges the location of the minima and maxima.

To investigate the VL ground states that emerge for different values of the anisotropy ratio, we performed MD simulations for $\kappa$ in the range $-6$ to 6.
The dynamics of vortex $i$ are governed by an overdamped equation of motion:
\begin{equation}
  \eta \frac{d\bm{r}_i}{dt} = \bm{F}^i_{vv} + \bm{F}^i_T.
\end{equation}
Here $\eta$ is the damping constant which is set equal to unity.
The force field from the surrounding vortices is given by
$\bm{F}_{vv} = -\bm{\nabla}(U) = (-\partial U/\partial x, -\partial U/\partial y)$, yielding
\begin{widetext}
  \begin{subequations}
    \begin{eqnarray}
      F_x/A_v
        & = & \cos (\theta) \, K_1(r) \left[ 1 + K_6 \cos^2 (3 \theta) + K_{12} \cos^2 (6 \theta) \right] 
              - \frac{\sin(\theta)}{r} K_0(r) \left[ 3 K_6 \sin (6 \theta) + 6 K_{12} \sin (12 \theta) \right] \\
      F_y/A_v
        & = & \sin (\theta) \, K_1(r) \left[ 1 + K_6 \cos^2 (3 \theta) + K_{12} \cos^2 (6 \theta) \right]
              + \frac{\cos(\theta)}{r} K_0(r) \left[ 3 K_6 \sin (6 \theta) + 6 K_{12}\sin (12 \theta) \right].
    \end{eqnarray}
  \end{subequations}
\end{widetext}
Thermal forces are modeled by Langevin kicks $\bm{F}^{i}_{T}$ with the properties $\langle \bm{F}_{T} \rangle = 0.0$ and
$\langle \bm{F}^i_T(t) \bm{F}^j_T(t') \rangle = 2 \eta k_B T \delta_{ij} \, \delta(t-t')$ where $k_B$ is the Boltzmann constant.
We perform simulated annealing by starting in a molten state with temperature $F^T = 6.0$ and gradually cooling the system to $F^T =0.0$.
The temperature is reduced by $\Delta F^T = -0.01$ every 40,000 simulation time steps, which is long enough to ensure that the system reaches an equilibrium state.

\section{Results}
Our results are organized as follows.
We first verify that the simulated annealing has converged properly and establish
the appropriate parameter regimes for investigation.
Next, we characterize the phase diagram for the VL rotation as a function of anisotropy.
Finally, we discuss the structure and energetics associated
with individual dislocations and defects as well as the domain formation process.

\subsection{Simulated annealing}
\label{Simulated annealing}
We simulate a two-dimensional system of size $L \times L$ with periodic boundary conditions in the $x$ and $y$ directions.
Distances are measured in units of the London penetration depth, $\lambda$.
To ensure that our results are not affected by the sample size, we performed simulations with different numbers of vortices $N_v$ while holding the vortex density constant at $\rho_v = N_v/L^2 = 0.4398/\lambda^2$.
We consider five system sizes:
$L = 36\lambda$ with 570 vortices,
$L = 72\lambda$ with 2,280 vortices,
$L = 108\lambda$ with 5,130 vortices,
$L = 144 \lambda$ with 9,120 vortices, and
$L = 180\lambda$ with 14,250 vortices.
We select values of $K_{12}= 0.001$, 0.005, 0.01, 0.0125 0.015, 0.0175, 0.02, 0.05, 0.1 and 0.2.
For each value of $K_{12}$, we vary the anisotropy ratio from $\kappa=-6$ to $\kappa=6$ in increments of 0.25 by modifying $K_6$.

Increasing or decreasing the values of $K_6$ and/or $K_{12}$ changes the magnitude of the energy potential experienced by each vortex, which is equivalent to a change in the effective vortex density.
To eliminate effects arising from a density difference, we define an effective magnetic field that is proportional to the two-dimensional integral of the interaction potential:
\begin{equation}
  B_{\rm{eff}} \propto \int_0^{\infty} r \, dr \int_0^{2\pi} d\theta \; U(r,\theta) \propto A_v \left( 1 + \frac{K_6}{2} + \frac{K_{12}}{2} \right).
  \label{Beff}
\end{equation}
Using $K_6 = 0$ and $A_v = 2.0$ as a reference, we set $A_v = 2 (2 + K_{12})/(2 + K_6 +K_{12})$ for each individual MD simulation, such that all simulations have the same value of $B_{\rm{eff}}$.

\begin{figure}
  \includegraphics[width=\columnwidth]{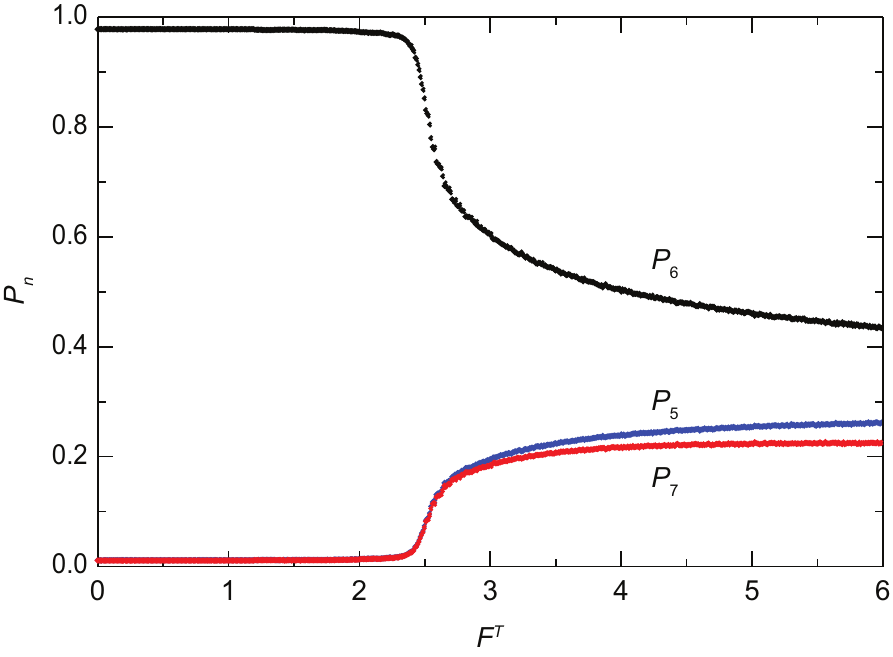}
    \caption{\label{Annealing}
      Coordination number $P_n$ vs temperature $F^T$ during an annealing process with $K_6 = 0.05$, $K_{12} = 0.05$, and a system size of $L = 144\lambda.$
      We plot only $P_5$, $P_6$, and $P_7$.
      The temperature decreases during the anneal, meaning that the system moves from right to left along the curves.
      At $F^T = 0$, we find $P_5 = 0.011$, $P_6 = 0.978$ and $P_7 = 0.011$.}
\end{figure}

The VL configuration is analyzed during the annealing process using a Voronoi polygon construction.
This yields the local coordination number $z_i$ of each vortex, which is used to compute the fractions
$P_n = \tfrac{1}{N_v} \sum_{i=1}^{N_v} \delta(z_i - n)$ for $n = 5$, 6, and 7.
Figure~\ref{Annealing} shows $P_5, P_6$ and $P_7$ versus $F^T$ obtained during a typical annealing process.
Initially, the system is in a high temperature molten state.
As $F^T$ is reduced, the vortices order into a triangular lattice with a few impurities at $F^T \sim 2.5$, as indicated by the increase in $P_6$ at this temperature.
We find qualitatively similar behavior during the annealing process for all choices of the parameters $A_v$, $K_6$ and $K_{12}$.
For the remainder of this paper we consider only the final VL configuration, obtained at $F^T = 0$.
\begin{figure*}
  \includegraphics[width=\textwidth]{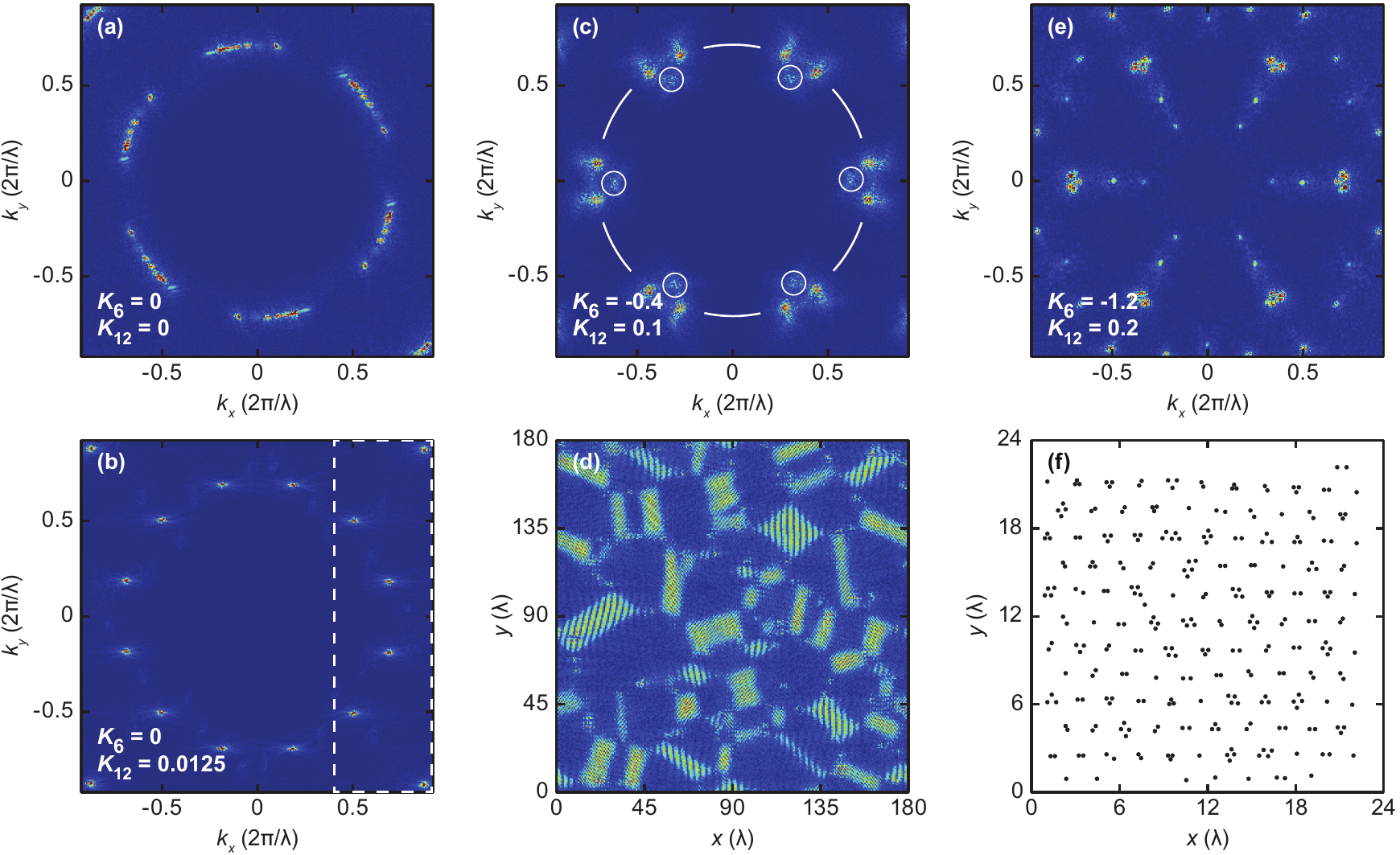}
    \caption{\label{AnisotropyLimits}
      Structure factor $S(\bm{k})$ plotted as a height field for
      (a) the isotropic case with $K_6 = K_{12} = 0$,
      (b) intermediate anisotropy with $K_6 = 0$ and $K_{12} = 0.0125$,
      (c) large anisotropy with $K_6=-0.4$ and $K_{12}=0.1$, and
      (e) very large anisotropy with $K_6=-1.2$ and $K_{12}=0.2$.
      The dashed rectangle in (b) indicates the reciprocal space area shown in Figs.~\ref{RotationTransition}(b)--\ref{RotationTransition}(h).
      Circle segments in (c) show the the expected radius ($0.71 \times 2\pi/\lambda$) for a uniform triangular VL with 14,250 vortices and a system size of $180\lambda \times 180\lambda$.
      (d) Inverse transform of the innermost contribution (indicated by the circles) to the structure factor in panel (c).
      (f) Real space vortex positions for the very large anisotropy system in (e), showing the formation of an ordered lattice of vortex clusters.}
\end{figure*}
To characterize the global VL configuration of the system after annealing we compute the structure factor,
$S(\bm{k}) \propto |\sum^{N_v}_i \exp(-i \bm{k} \cdot \bm{r}_i)|^2$.
To achieve meaningful results from the simulations it is necessary to establish the appropriate range of values for $K_6$ and $K_{12}$.
If the anisotropy amplitudes are too small they will not significantly affect the VL, and the results will be the same as for an isotropic system.
In addition, due to the weaker anisotropy, boundary effects become more significant, often dictating the orientation of the VL.
Conversely, values that are too large may lead to instabilities, such as the previously observed vortex chain states in the case of a single four-fold anisotropy \cite{Olszewski:2018fp}.

In Fig.~\ref{AnisotropyLimits} we illustrate representative examples of the structure factor for low, intermediate, and high anisotropy samples.
For the isotropic $K_6 = K_{12} = 0$ case of Fig.~\ref{AnisotropyLimits}(a) we observe an almost complete ``powder ring'' arising from a number of randomly oriented VL domains.
Here, the gaps in the ring are due to the finite system size.
The radius of the ring agrees with the calculated value for a triangular lattice, $k_0 = (2/\sqrt{3})^{1/2} \; 2\pi (\sqrt{14250}/180\lambda) = 0.71 (2\pi/\lambda)$.
The narrow radial width of $S(\bm{k})$ corresponds to a very uniform vortex spacing, indicating that the individual domains are well ordered.
In the intermediate anisotropy sample with $K_6 = 0$ and $K_{12} = 0.0125$ in Fig.~\ref{AnisotropyLimits}(b), the intensity of $S(\bm{k})$ still lies on a circle with the same radius as before, but it is now concentrated in 12 sharp (Bragg) peaks.
This indicates the presence of VL domains that are oriented along one of two different directions separated by $30^{\circ}$, corresponding to the minima in the interaction potential in Fig.~\ref{Potentials}(d).

When the anisotropy is large, as in Fig.~\ref{AnisotropyLimits}(c) at $K_6 = -0.4$ and $K_{12} = 0.1$, the maxima in $S(\bm{k})$ broaden significantly in the radial direction and develop an internal structure.
Figure~\ref{AnisotropyLimits}(d) shows the inverse transform, $|\int S(\bm{k}) \, \exp(i \bm{k} \cdot \bm{r}) \, d^2\bm{k}|^2$, for this system, calculated using only the innermost (lowest $k$) contributions to the structure factor.
It reveals a non-uniform vortex spacing, with intertwined regions of low (bright) and high (dark) vortex density. 
The heterogeneous vortex density is due to an instability caused by the deep minima that appear in the interaction potential for high values of the anisotropy.
A more extreme case is shown in Fig.~\ref{AnisotropyLimits}(e) for a very large anisotropy of $K_6 = -1.2$ and $K_{12} = 0.2$.
Here additional peaks emerge in $S(\bm{k})$, indicating the development of a superstructure.
This is illustrated directly in the real space image of Fig.~\ref{AnisotropyLimits}(f), where closely bound clusters, each containing two to five vortices, are arranged in a periodic lattice.
Here the large-scale six-fold ordering is produced by the strong six-fold anisotropy term. 
This is similar to the cluster crystals found in particle-based systems with competing isotropic interactions\cite{Reichhardt10,Meng17}.
It may be possible that the true lowest energy state would contain clusters of a single size, such as three vortices.
The competition between the two anisotropies could frustrate the system, however, producing a variety of ground states with roughly the same energy that prevent the system from crystallizing on the cluster level and giving rise to the strong dispersion in the cluster size.

\subsection{Rotational transition}
\label{rottrans}
\begin{figure*}
  \includegraphics{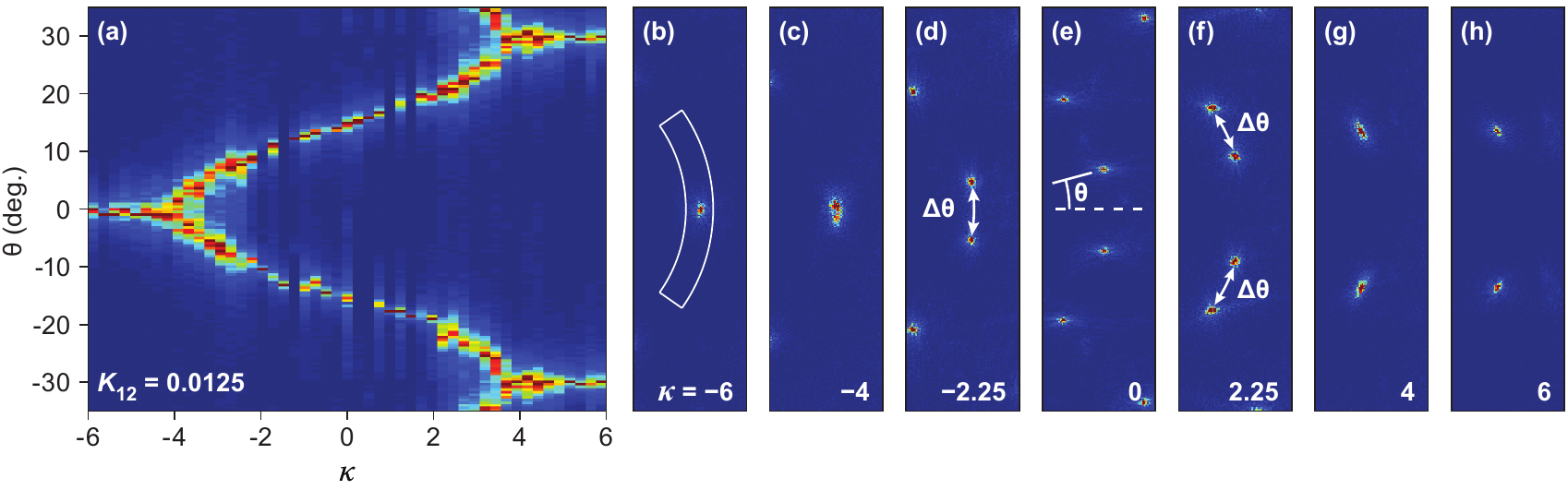}
    \caption{\label{RotationTransition}
      Vortex lattice rotation transition for $K_{12} = 0.0125$.
      (a) Azimuthal dependence of the structure factor, averaged over all six $60^{\circ}$ segments, vs the anisotropy ratio $\kappa$.
      The azimuthal angle range is illustrated in panel (b).
      (b)--(h) $S(\bm{k})$ for select values of $\kappa$ for the region of reciprocal space indicated in Fig.~\ref{AnisotropyLimits}(b).
      Here $\kappa=$ (b) -6, (c) -4, (d) -2.25, (e) 0, (f) 2.25, (g) 4, and (h) 6.
      The definition of the VL splitting angle $\Delta \theta$ for $\kappa \leq 0$ and $\kappa \geq 0$ is shown in panels (d) and (f) respectively, while the rotation angle $\theta$ is shown in panel (e).}
\end{figure*}
As discussed in Sect.~\ref{Methods}, changing the anisotropy ratio $\kappa$ causes the minima in the interaction potential to both split and rotate within the range $\theta = [-30^{\circ},30^{\circ}]$.
Figure~\ref{RotationTransition}(a) illustrates the azimuthal dependence of the structure factor as $\kappa$ is varied from $\kappa = -6$ to $\kappa = +6$ in steps of $\Delta\kappa=0.25$ for samples with fixed $K_{12} = 0.0125$ and $L = 180\lambda$.
Due to the six-fold symmetry that appears both in the interaction potential in Fig.~\ref{Potentials} and in the structure factor plots in Fig.~\ref{AnisotropyLimits}, to obtain the intensity in Fig.~\ref{RotationTransition}(a), we take the average of six $60^{\circ}$ segments.
For $\kappa < -4$ we find a single peak at $\theta  = 0^{\circ}$, indicating the presence of a single preferred domain orientation.
When $\kappa$ increases above $\kappa = -4$, the peak splits in two, with a peak separation that grows with increasing $\kappa$.
Finally, for $\kappa > 4$, the split peaks recombine, with the peak at $\theta = \pm 30^{\circ}$ merging with the peak arising from the neighboring $60^{\circ}$ segment.
This behavior indicates that when $|\kappa| < 4$, the VL breaks into domains that rotate either clockwise or counterclockwise between the two high-symmetry directions, while when $|\kappa| > 4$, there is a single preferred direction of the VL.
The series is symmetric around $\kappa = 0$, with identical behavior appearing under a reflection and a $30^{\circ}$ shift in $\theta$.
Figures~\ref{RotationTransition}(b)--~\ref{RotationTransition}(h) show the structure factor for select values of $\kappa$ along the rotation transition.
In the vicinity of the rotation onset at $\kappa = \pm 4$, the structure factor broadens in the azimuthal direction, as shown in Fig.~\ref{RotationTransition}(c) for $\kappa = +4$ and Fig.~\ref{RotationTransition}(g) for $\kappa = -4$, before splitting into two well-defined peaks in Figs.~\ref{RotationTransition}(d) through \ref{RotationTransition}(f) at $\kappa = -2.25$, 0, and 2.25, respectively.
The broadening for $|\kappa| \sim 4$ is likely due to the shallow minima that emerge in the interaction potential, illustrated in Figs.~\ref{Potentials}(b) and \ref{Potentials}(f).

\begin{figure}
  \includegraphics[width=\columnwidth]{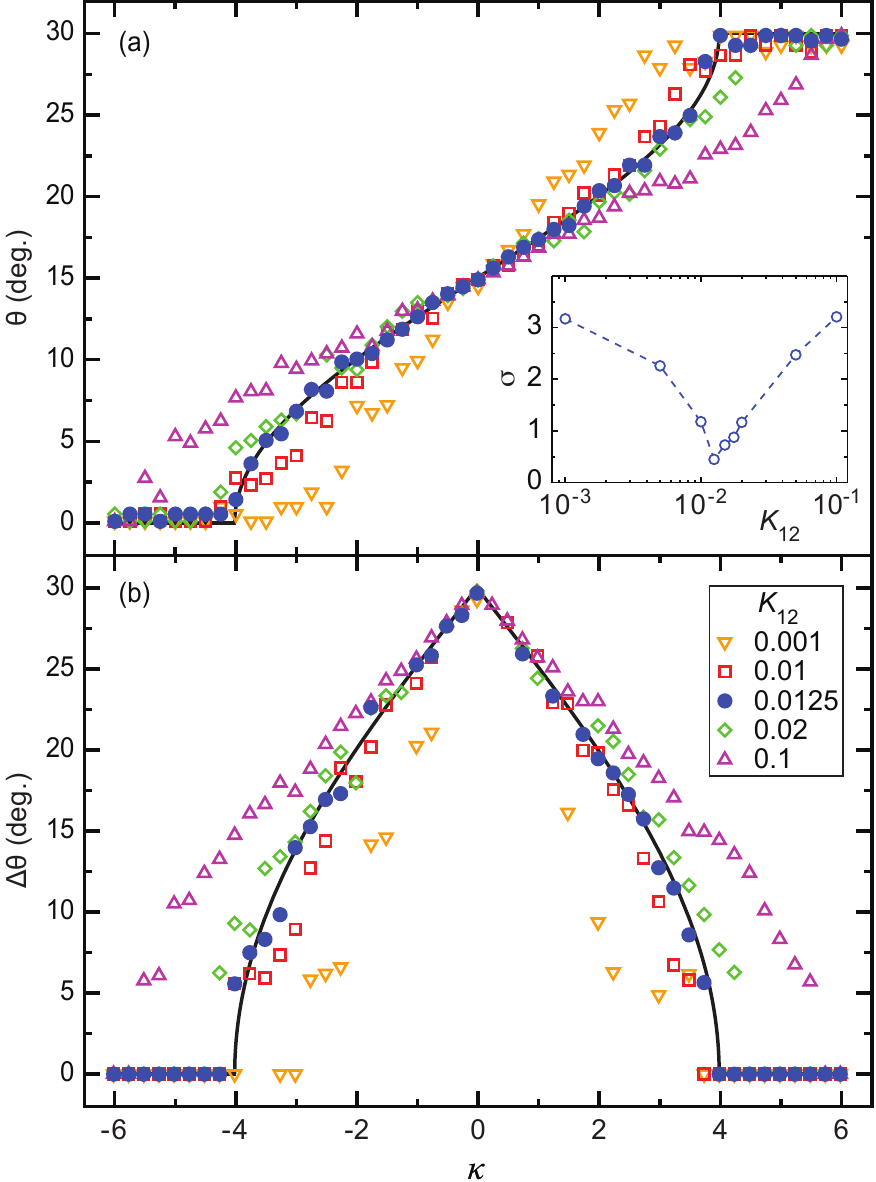}
    \caption{\label{Splitting}
      (a) Peak position $\theta$ and (b) peak splitting $\Delta \theta$, obtained from the structure factor, plotted vs anisotropy ratio $\kappa$ for different values of $K_{12}$.
      Lines indicate the locations of the minima in the interaction potential, Eq.~(\ref{thetamin}).
      The inset in (a) shows the standard deviation $\sigma$ between the simulated value of $\theta$ and the actual location of the potential minimum as a function of $K_{12}$.}
\end{figure}

The VL rotation transition can be quantified by either the angle between domains, given by the structure factor peak splitting $\Delta \theta$ indicated in Figs.~\ref{RotationTransition}(d) and \ref{RotationTransition}(f), or by the rotation $\theta$ of one of the domains relative to a fixed direction, as shown in Fig.~\ref{RotationTransition}(e).
Figure~\ref{Splitting} shows both the VL domain splitting $\Delta \theta$ and rotation $\theta$ versus $\kappa$ for different values of $K_{12}$ and a system size of $L = 180\lambda.$
Also shown for comparison are the location $\theta_{\rm min}$ and separation $\Delta \theta$ of the minima in the interaction potential given by Eq.~(\ref{thetamin}), corresponding to the lowest energy configuration for a single VL domain.
Here, $\Delta\theta = 2 \theta_{\text{min}}$ for $\kappa \leq 0$ and $\Delta\theta = 60^{\circ} - 2 \theta_{\text{min}}$ for $\kappa \geq 0$.
In the absence of effects arising due to the presence of grain boundaries, the MD simulation results are expected to follow these calculated curves, and we find good agreement for values of $K_{12}$ in the range $K_{12} = 0.01$ to 0.02.
For values outside this range the simulation results deviate from the curves, particularly in the vicinity of $\kappa = \pm 4$.
For $K_{12} < 0.01$, the transition from $\Delta\theta = 0^\circ$ to $\Delta\theta = 30^\circ$ occurs over a narrower range of the anisotropy $\kappa$, most likely because the minima that appear in the interaction potential near $\kappa = \pm 4$ are too shallow to produce ordered VL domains when the splitting is small.
For these small values of $K_{12}$, we posit that the effects of the square sample geometry are proportionally stronger and change the nature of the observed transition.
Conversely, for $K_{12} > 0.02$, the splitting process continues beyond $\kappa=\pm 4$, which we attribute to the pattern-forming instability discussed in Sect.~\ref{Simulated annealing} and illustrated in Figs.~\ref{AnisotropyLimits}(c)--\ref{AnisotropyLimits}(f).
We obtain the best agreement between Eq.~(\ref{thetamin}) and the MD simulation results when $K_{12} = 0.0125$, as measured by the deviation $\sigma = \sqrt{\sum (\theta_{\text{MD}} - \theta_{\text{min}})^2/(M-1)}$ between the simulation results and the theory, where $M = 49$ is the number of simulations in a single sequence.
As shown in the inset to  Fig.~\ref{Splitting}(a), $\sigma$ is minimized at $K_{12}=0.0125$.
The relative narrowness of the minimum in $\sigma$ as a function of $K_{12}$ underscores the importance of optimizing the parameters used for the MD simulations.
It is possible that a different choice of interaction potential that avoids the wide and flat minima for $|\kappa| \approx 4$ would have a broader minimum in $\sigma$ and thus require less optimization.
\begin{figure}
  \includegraphics[width=\columnwidth]{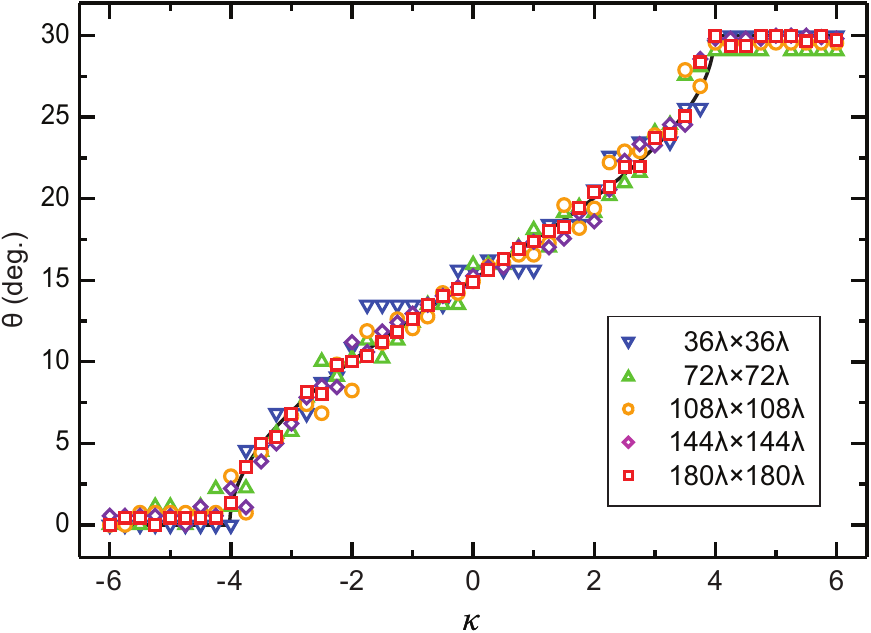}
    \caption{\label{SystemSize}
      Structure factor peak position $\theta$ vs anisotropy ratio $\kappa$ for different system sizes at $K_{12} = 0.0125$.
      The line corresponds to the interaction potential minimum $\theta_{\rm min}$ obtained from Eq.~(\ref{thetamin}).}
\end{figure}
\begin{figure*}
  \includegraphics{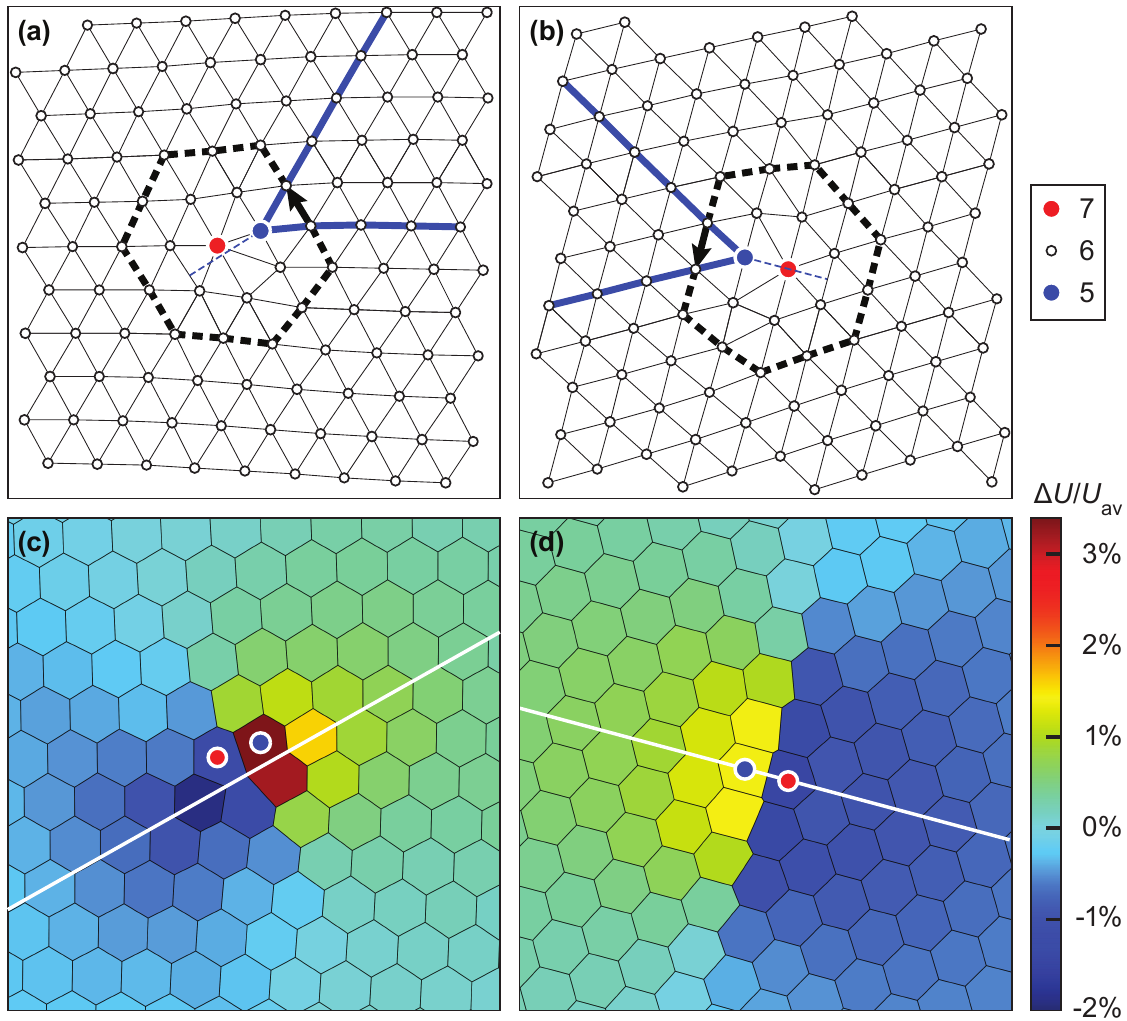}
    \caption{\label{EdgeDislocations}
      Edge dislocations observed in (a,c) the six-fold anisotropy regime ($\kappa = -5.25$) and (b,d) the 12-fold anisotropy regime ($\kappa = 0$).
      Delaunay triangulations (a,b) indicate the Burgers circuit (dashed black line), non-zero Burgers vector (black arrow), the extra lattice planes originating on the five-coordinated vortex (solid blue lines) and the direction of their bisector (dashed blue line).
      Five-, six- and seven-coordinated vortices are plotted as blue, white, and red circles, respectively.
      The corresponding heatmaps (c,d) show the deviation $\Delta U/U_{\text{av}}$ of the energy $U$ of each vortex, based on its interactions with the surrounding VL, from the system wide average energy $U_{\text{av}}$.
      White lines indicate high symmetry directions used to determine the range of the dislocations.}
\end{figure*}

While the evolution of the rotation transition obtained from the MD simulations depends sensitively on the anisotropy amplitude as discussed above, it is independent of the system size.
This is illustrated in Fig.~\ref{SystemSize}, where results from samples with $L$ ranging from $L = 36\lambda$ to $L = 180\lambda$ agree within stochastic fluctuations.
In all cases, $\sigma < 1$.
As the number of vortices is reduced, the fraction of simulations that terminate with a single VL domain increases.
In such cases it is not possible to define a value of the peak splitting $\Delta \theta$, and therefore we plot only the VL rotation $\theta$.

\subsection{Dislocations and Defects}
At the local scale the annealed lattices frequently contain vortices that are not six-fold coordinated, even though the structure factor shows sharp peaks as in Figs.~\ref{AnisotropyLimits}(b) and \ref{RotationTransition}.
In the following we characterize the observed lattice imperfections occurring within the optimal range $0.01 \geq K_{12} \geq 0.02$ determined above.
We compare systems with $|\kappa| > 4$, corresponding to a six-fold anisotropy in the interaction potential in Fig.~\ref{Potentials}, to systems with $|\kappa| \leq 2$, corresponding to a well developed 12-fold interaction potential anisotropy.

The elemental lattice imperfection that we observe is a 5-7 edge dislocation, shown in Fig.~\ref{EdgeDislocations}(a) for a sample with $\kappa = -5.25$ in the six-fold anisotropy regime and in Fig.~\ref{EdgeDislocations}(b) for a sample with $\kappa = 0$ in the 12-fold anisotropy regime.
This dislocation inserts two new lattice half-planes originating on the five-coordinated vortex and separated by $60^{\circ}$,
producing a non-zero Burgers vector\cite{AshcroftMerminTinkham:1976} (open Burgers circuit) as highlighted in Fig.~\ref{EdgeDislocations}(a,b).
In the 12-fold anisotropy regime, the vector bisecting the two new lattice planes runs along the line connecting the 5-7 vortex pair.
This high degree of symmetry is possible due to the $30^\circ$ separation between the minima of the interaction potential.
In the six-fold anisotropy regime this symmetry is broken and, as shown in Fig.~\ref{EdgeDislocations}(a), the vortex with seven-fold coordination is rotated away from the bisector direction, as the latter now passes through a maximum in the potential energy landscape.
We note that ``free'' edge dislocations, {\it i.e.} 5-7 vortex pairs that are not located along VL grain boundaries described in Sect.~\ref{GB}, do not appear for intermediate values $2 < |\kappa| < 4$ 
of the anisotropy ratio; they form only in the six-fold and 12-fold dominated regimes defined above.

The difference between the six-and 12-fold anisotropy is visible in the distribution of vortex energies near the edge dislocation, shown in Figs.~\ref{EdgeDislocations}(c) and \ref{EdgeDislocations}(d).
The energy $U_i$ of vortex $i$ is given by $U_i = \sum_{j \neq i} U(r_{ij},\theta_{ij})$, where $U(r_{ij},\theta_{ij})$ is defined in Eq.~(\ref{K612potential}).
Here $r_{ij}$ and $\theta_{ij}$ are, respectively, the distance and angle to the horizontal axis of the vector joining
vortices $i$ and $j$.
We plot the relative energy difference $\Delta U/U_{\text{av}}$, where $\Delta U = U_i - U_{\text{av}}$ and
$U_{\text{av}}=N_v^{-1}\sum_i^{N_v}U_i$ is the average vortex energy of the system.
In all cases, $U_{\text{av}}$ is close to the value $1.678 A_v$ obtained from Eq.~(\ref{K612potential}) for an isotropic interaction potential.
Both the six- and 12-fold regimes show the formation of an energy ``dipole'' due to the compression (tension) experienced by the vortices on the same (opposite) side as the new lattice planes.
The rotation of the seven-fold coordinated vortex away from the bisector in the six-fold regime produces larger deviations from the average energy compared to the 12-fold regime, where the 5-7 vortex pair remains aligned with the bisector.
On the other hand, the disturbance in the energy profile produced by the 5-7 vortex pair extends further in the direction perpendicular to the bisector for the 12-fold regime than in the six-fold regime, meaning that the energy is more localized for the six-fold regime and less localized for the 12-fold regime.
In both cases, the vortex energies are highly symmetric around the bisector direction.
These results indicate that focusing solely on the vortex coordination number, which in the six-fold regime is asymmetric around the bisector, can be misleading.
This is because small displacements can change which of the vortices near the edge dislocation are five- or seven-coordinated, without significantly altering their energies.

\begin{figure}
  \includegraphics[width=\columnwidth]{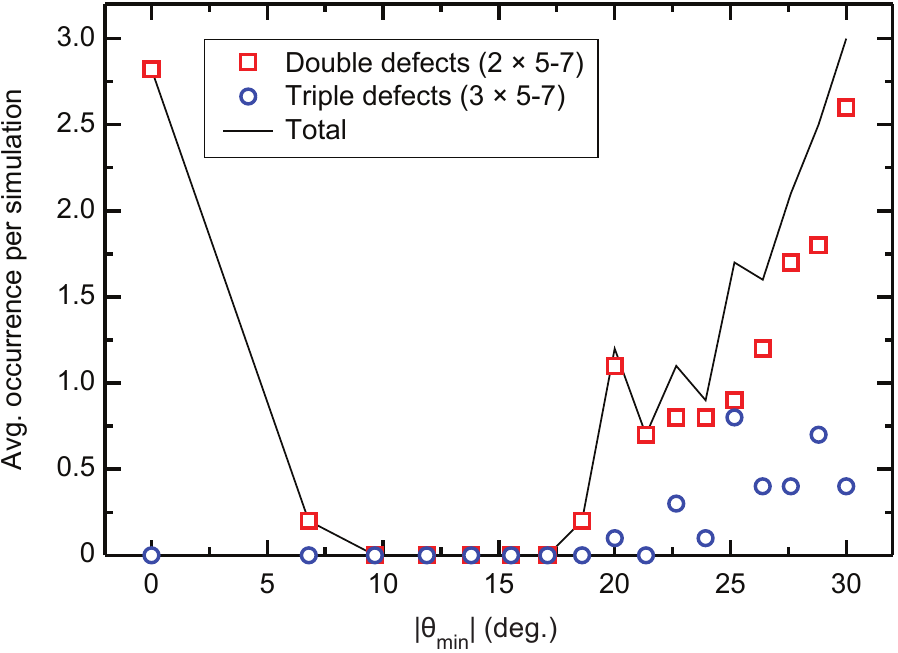}
    \caption{\label{DefectCount}
      Average number of defects occurring per simulation vs the location $|\theta_{\rm min}|$ of the minimum in the interaction potential.
      Double and triple defects correspond to combinations of two or three 5-7 dislocations, respectively.}
\end{figure}

By merging two or more 5-7 edge dislocations, it is possible to create localized defects with a closed Burgers circuit (zero Burgers vector) and an average vortex coordination number equal to six.
Defects emerge spontaneously when the vortex-vortex interactions are anisotropic, but must be ``seeded'' into equilibrated isotropic vortex lattices by either adding or removing a particle and the letting the remaining system relax.\cite{Jain:2000fh,Libal:2007tv}
Many of the defects we find were previously reported in other two-dimensional systems including colloids,\cite{Pertsinidis:2001hz,Libal:2007tv} skyrmions \cite{Pollath17,Nakajima17a,Huang18}, graphene \cite{Warner12} and silica bi-layers \cite{Bjorkman13}, and throughout this paper we adopt and expand the nomenclature used in Refs.~\onlinecite{Jain:2000fh,Pertsinidis:2001hz,Libal:2007tv}.
As was the case for free edge dislocations, both vacancy and interstitial defects appear almost exclusively in the six- and 12-fold regimes, with only $\sim\!1\%$ of defect observations occurring for $2 < |\kappa| < 4$.
This is illustrated in Fig.~\ref{DefectCount}, where we plot the average defect count per simulation as a function of the location $|\theta_{\rm min}|$ of the minimum in the interaction potential.
\begin{figure*}
  \includegraphics{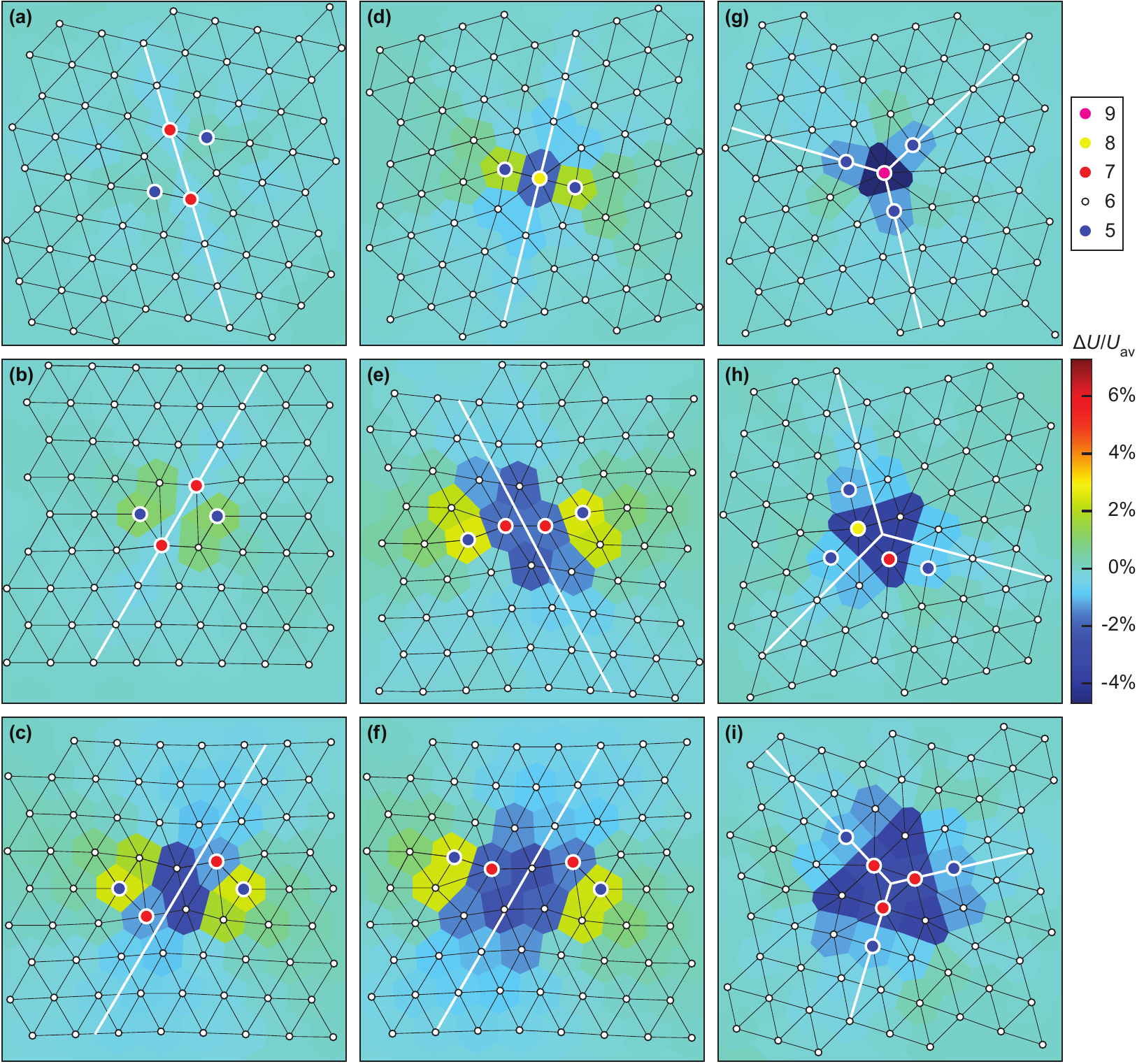}
    \caption{\label{Vacancies}
      Vacancy defects.
      Delaunay triangulations overlaid with heatmaps showing respectively the coordination and energy for each vortex.
      Two-fold symmetric crushed:
      (a) $V_{2b}^{(12)}$, 
      (b) $V_{2a}^{(6)}$, 
      (c) $D_{2b}^{(6)}$. 
      Two-fold symmetric split:
      (d) $SV^{(12)}$, 
      (e) $SD_a^{(6)}$, 
      (f) $STr_a^{(6)}$. 
      Three-fold symmetric:
      (g) $D_3^{(12)}$, 
      (h) $Tr_3^{(12)}$, 
      (i) $Te_3^{(12)}$. 
      White lines indicate high symmetry directions used to determine the range of vacancy defects.}
\end{figure*}
\begin{figure*}
  \includegraphics{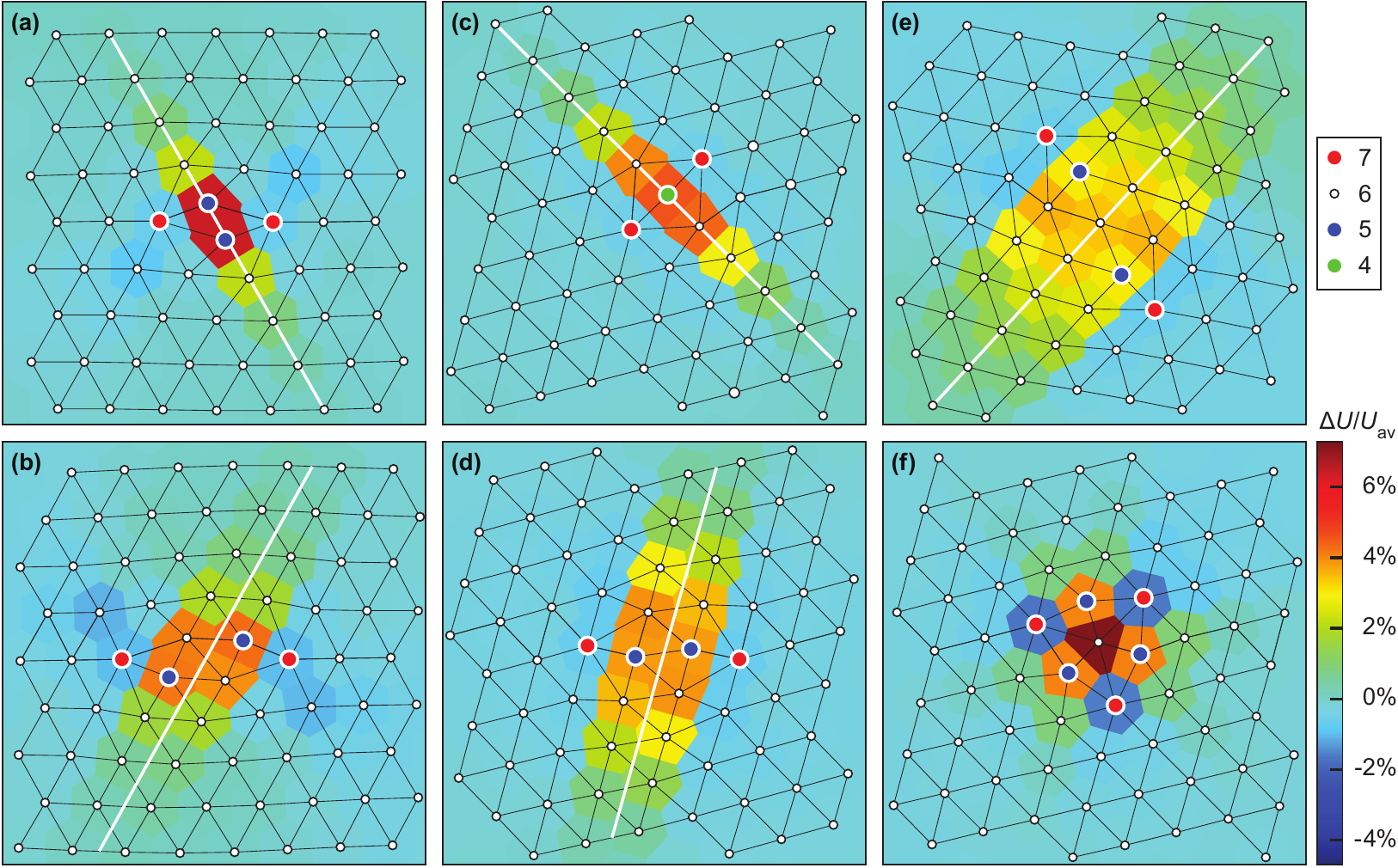}
    \caption{\label{Interstitials}
      Interstitial defects.
      Delaunay triangulations overlaid with heatmaps showing respectively the coordination and energy for each vortex.
      Two-fold symmetric crushed:
      (a) $I_2^{(6)}$, 
      (b) $I_{2d}^{(6)}$. 
      Two-fold symmetric split:
      (c) $SI_{2}^{(12)}$, 
      (d) $SDI_{2}^{(12)}$, 
      (e) $STrI_{2d}^{(12)}$. 
      Three-fold symmetric:
      (f) $I_3^{(12)}$. 
      White lines indicate high symmetry directions used to determine the range of interstitial defects.}
\end{figure*}
Here the six-fold regime corresponds to $\theta_{\text{min}} = 0$ and the 12-fold regime to $|\theta_{\text{min}}| \geq 20^{\circ}$.
As we would expect, defects created by combining two 5-7 dislocations appear more frequently than those that combine three.

In Fig.~\ref{Vacancies}, we provide a survey of the typical vacancy defects that we observe.
Figures~\ref{Vacancies}(a) and \ref{Vacancies}(b) show examples of single (one missing vortex) two-fold symmetric crushed vacancies which we label $V_{2a/2b}^{(x)}$, where the superscript $x$ denotes the symmetry of the interaction potential (six- or 12-fold).
These are the most common vacancy defects, constituting $\sim\!87\%$ of observed vacancies in the six-fold regime and
$\sim\!34\%$ of observed vacancies in the 12-fold regime.
The energy $\Delta U/U_{\text{av}}$ is almost featureless in the 12-fold regime, but shows some deviations from the average vortex energy in the six-fold regime.
Larger excursions from the average energy appear in the disjoint symmetric crushed divacancy $D_{2b}^{(6)}$ shown in Fig.~\ref{Vacancies}(c), which occurs only in the six-fold regime (constituting $\sim\!12\%$ of observations).
The two-fold symmetric split vacancies in Figs.~\ref{Vacancies}(d)--\ref{Vacancies}(f) display a trend similar to the crushed vacancies, with deviations from the average energy increasing upon passing from the single vacancy $SV^{(12)}$ to the divacancy $SD_a^{(6)}$ and finally to the disjoint trivacancy $STr_a^{(6)}$.
Here, $SV^{(x)}$ and $SD_a^{(x)}$ occur almost exclusively in the 12-fold regime, where together they account for $\sim\!17\%$ of the observations, and we found only a single occurrence of $STr_a^{(x)}$ for each of the two symmetries throughout all of the simulations.
The three-fold symmetric vacancies in Figs.~\ref{Vacancies}(g)--\ref{Vacancies}(i) correspond to a divacancy $D_3^{(12)}$, a trivacancy $Tr_3^{(12)}$, and a tetravacancy $Te_3^{(12)}$, respectively, and appear only in the 12-fold regime.
In contrast to the two-fold symmetric vacancies, the three-fold symmetric vacancies contain no vortices with $U_i > U_{\text{av}}$.
The $Tr_3^{(12)}$ vacancy is an additional example of a defect where the energy distribution reveals a higher degree of symmetry than what is indicated by the Delaunay triangulation.
Finally, we note that in the two-fold symmetric crushed and split vacancy defects shown in Figs.~\ref{Vacancies}(c) and \ref{Vacancies}(f), it is in principle possible to remove an arbitrary number of vortices within a lattice plane by increasing the separation between the 5-7 dislocations along the direction joining the five-fold coordinated vortices.
In contrast, separating the 5-7 dislocations along a line oriented 60$^{\circ}$ from this direction does not create additional vacancies.
In the interest of brevity we will consider such cases as separate, closely spaced dislocations rather than localized defects.

In Figure~\ref{Interstitials} we show typical examples of interstitial defects.
In analogy with the vacancies discussed above, we observe two-fold symmetric crushed interstitials $I_2^{(6)}$ [Fig.~\ref{Interstitials}(a)] and their disjoint variants $I_{2d}^{(6)}$ [Fig.~\ref{Interstitials}(b)], where the latter is a double interstitial.
In the six-fold regime, nearly all observed defects ($\sim\!98\%$) are of $I_2^{(6)}$ or $I_{2d}^{(6)}$ type, while the $I_2^{(12)}$ and $I_{2d}^{(12)}$ defects constitute $\sim\!55\%$ of observations in the 12-fold regime.
We also find twofold symmetric split interstitials in the 12-fold regime that incorporate one ($SI_{2}^{(12)}$), two ($SDI_{2}^{(12)}$) or three ($STrI_{2d}^{(12)}$) additional vortices, as shown in Figs.~\ref{Interstitials}(c)--\ref{Interstitials}(e).
In contrast to the behavior found for vacancies, the deviation from the average vortex energy for the interstitial defects decreases upon passing from $SI_{2}^{(12)}$ to $STrI_{2d}^{(12)}$.
We observed only a single instance of a three-fold symmetric single interstitial defect $I_3^{(12)}$, shown in Fig.~\ref{Interstitials}(f).
Unlike the three-fold symmetric vacancies, this interstitial defect includes vortices with energies both below and above $U_{\text{av}}$.
As with vacancy defects, the separation between the 5-7 dislocations within the two-fold symmetric crushed and split interstitials in Figs.~\ref{Interstitials}(b) and \ref{Interstitials}(e) can be increased in order to insert a finite lattice plane containing an arbitrary number of vortices.

\begin{figure*}
  \includegraphics[width = 2 \columnwidth]{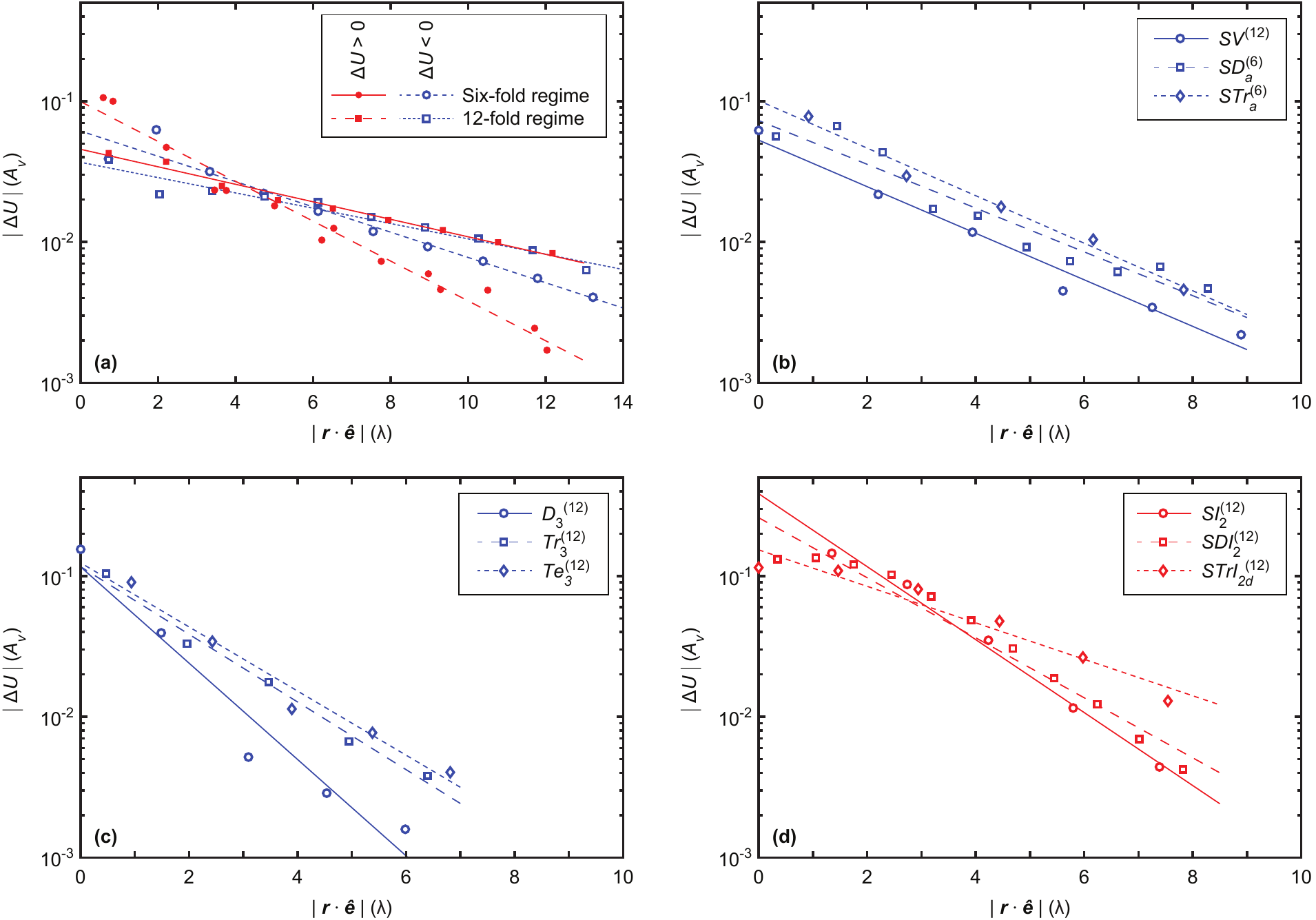}
    \caption{\label{Range}
      Range of the energy deviation of dislocations and defects, plotted as $|\Delta U|$ vs $|{\bm r} \cdot \hat{\bm{e}}|$, a distance taken along the high-symmetry direction indicated by white lines in Figs.~\ref{EdgeDislocations}, \ref{Vacancies}, and \ref{Interstitials}.
      In all cases lines are exponential fits to the data.
      (a) Edge dislocations.
      (b) Double vacancy defects.
      (c) Triple vacancy defects.
      (d) Interstitial defects.}
\end{figure*}

Considering the defect energetics, it is not surprising that the $SV^{(12)}$, $D_3^{(12)}$, $Tr_3^{(12)}$ and $Te_3^{(12)}$ vacancies as well as the $SI_{2}^{(12)}$, $SDI_{2}^{(12)}$, $STrI_{2d}^{(12)}$ and $I_3^{(12)}$ interstitials appears almost exclusively in the 12-fold anisotropy regime.
As shown in Figs.~\ref{Vacancies} and \ref{Interstitials}, they all contain vortex pairs that are rotated by $30^{\circ}$ relatively to the surrounding VL planes.
In the 12-fold anisotropy regime, this corresponds to the location of an additional minimum in the interaction potential, as discussed above for the edge dislocations, making such pairs energetically favorable and stabilizing the defect.
These additional minima in the interaction potential are absent in the six-fold anisotropy regime, causing the above-mentioned defects to be very energetically costly.
To further characterize the dislocations and defects, we consider how the vortex energy $U$ approaches $U_{\text{av}}$ along the high symmetry directions indicated by white lines in Figs.~\ref{EdgeDislocations}, \ref{Vacancies} and \ref{Interstitials}.
We plot the spatial variation of $\Delta U = U - U_{\text{av}}$ versus the parallel component of the radial distance $| \bm{r} \cdot \hat{\bm{e}}|$ in Fig.~\ref{Range}, where $\bm{r}$ is the vortex position relative to the center of the dislocation/defect and $\hat{\bm{e}}$ is a unit vector along the relevant high symmetry direction.
We include only a selection of representative defects in Figs.~\ref{Range}(b)-\ref{Range}(d) for clarity.
Motivated by the near exponential decay of $K_0(r)$ in the interaction potential in Eq.~(\ref{K612potential}) for $r \gg 1$,
we fit $|\Delta U|$ by $\exp[-| \bm{r} \cdot \hat{\bm{e}}|/a]$ to obtain the range $a$ for each dislocation and defect.  
Table~\ref{CharLength} lists the values of $a$ obtained in this way.
The only omissions are $V_{2b}^{(12)}$ and $V_{2a}^{(6)}$, where the energy variation is insufficient to yield a reliable fit, and $I_3^{(12)}$ which has both positive and negative deviations from $U_{\text{av}}$.
All values of $a$ exceed the range $a \approx 1.1 \lambda$ obtained directly from $K_0(r)$ for an isolated vortex.

The 5-7 edge dislocations have the largest ranges, and we find notable range differences between the six- and 12-fold regimes.
As shown in Fig.~\ref{Range}(a), the 12-fold regime has lower values of $|\Delta U|$ and an essentially symmetric decay of the energy along the symmetry line on either side of the defect center, but the defect ranges are greater than what we observe in the six-fold regime.
In contrast, the decay of $|\Delta U|$ in the six-fold regime is different on the positive and negative sides, most likely due the lower degree of symmetry discussed previously.
All two-fold symmetric vacancies have roughly the same range $a \approx 2.6 \lambda$, while the three-fold symmetric vacancy ranges are shorter, although the range increases on passing from $D_3^{(12)}$ to $Te_3^{(12)}$.
Thus all vacancy defects have a shorter range than the 5-7 edge dislocations, likely due to the closed Burgers circuit and corresponding absence of additional lattice planes.
For interstitial defects we find a greater variation in the range, but again the range is always smaller than for the 5-7 edge dislocations when comparing the six- and 12-fold regimes separately.
\begin{table}
  \caption{\label{CharLength}
    Characteristic lengths for lattice imperfections in units of the penetration depth.}
    \begin{ruledtabular}
      \begin{tabular}{l|cc}
          & $U - U_{\text{av}} > 0$ & $U - U_{\text{av}} < 0$ \\ \hline
        $5$-$7^{(6)}$      & $3.1 \pm 0.3$ & $4.8 \pm 0.8$ \\
        $5$-$7^{(12)}$     & $7.0 \pm 0.7$ & $8.0 \pm 1.5$ \\ \hline
        $D_{2b}^{(6)}$     &               & $2.4 \pm 0.4$ \\ 
        $SV^{(12)}$        &               & $2.6 \pm 0.6$ \\ 
        $SD_a^{(6)}$       &               & $2.8 \pm 0.7$ \\ 
        $STr_a^{(6)}$      &               & $2.6 \pm 0.5$ \\ 
        $D_3^{(12)}$       &               & $1.3 \pm 0.5$ \\ 
        $Tr_3^{(12)}$      &               & $1.8 \pm 0.4$\\ 
        $Te_3^{(12)}$      &               & $1.9 \pm 0.6$ \\ \hline 
        $I_2^{(12)}$       & $1.9 \pm 0.3$ &     \\ 
        $I_2^{(6)}$        & $1.6 \pm 0.2$ &     \\ 
        $I_{2d}^{(6)}$     & $2.1 \pm 0.2$ &     \\ 
        $SI_{2}^{(12)}$    & $1.7 \pm 0.3$ &     \\ 
        $SDI_{2}^{(12)}$   & $2.0 \pm 0.3$ &     \\ 
        $STrI_{2d}^{(12)}$ & $3.4 \pm 1.2$ &     \\ 
      \end{tabular}
    \end{ruledtabular}
\end{table}

\subsection{Grain Boundaries}
\label{GB}
\begin{figure*}
  \includegraphics{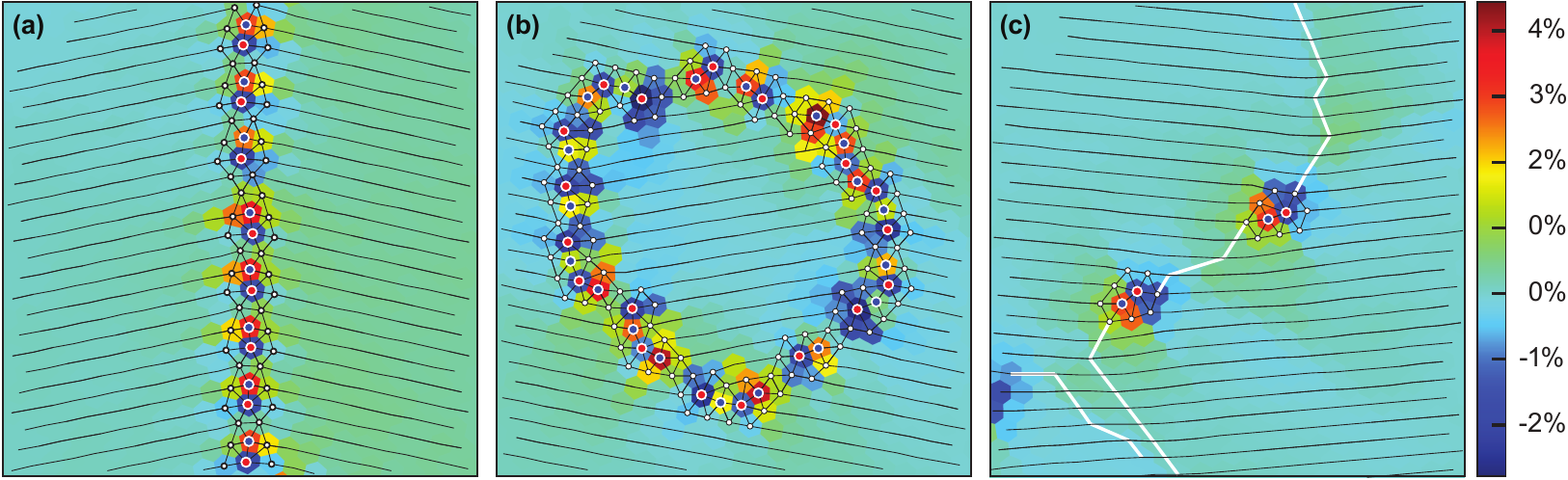}
    \caption{\label{GrainBoundaries}
      Grain boundaries plotted as Delaunay triangulations for different values of the anisotropy ratio: (a) $\kappa = -1.75$, $\Delta \theta = 21.4^{\circ}$, (b) $\kappa = -2.00$, $\Delta \theta = 20.0^{\circ}$ and (c) $\kappa = -3.25$, $\Delta \theta = 11.9^{\circ}$.
      For simplicity, we plot only five- and seven-fold coordinated vortices along with their adjacent six-fold coordinated neighbors, and the VL orientation in the bulk is indicated by a single set of  lattice planes.
      Vortices are colored according to their coordination number as in Fig.~\ref{EdgeDislocations}, and the plots are overlaid on a heatmap indicating the relative energy $U/U_{\text{av}}$ for each vortex.
      In panel (c), the white line indicates the location of the grain boundary.}
\end{figure*}

The presence of two minima in the interaction potential within each $60^{\circ}$ angular segment when $-4 < \kappa < 4$ produces a two-fold degeneracy for the triangular VL.
This two-fold degeneracy leads to the formation of VL domains that are rotated by an angle $\Delta \theta$ with respect to each other, as indicated by the twelve peaks in the structure factor in Fig.~\ref{RotationTransition}(d-f); however, due to the finite size of our system, not all simulations result in the formation of a multi-domain state due.
In Fig.~\ref{GrainBoundaries} we show three representative examples of grain boundaries (GB) separating VL domains.
Fig.~\ref{GrainBoundaries}(a) shows a straight continuous GB  at $\kappa = -1.75$, Fig.~\ref{GrainBoundaries}(b) illustrates a curved continuous GB at $\kappa = -2.0$, and Fig.~\ref{GrainBoundaries}(c) shows a meandering GB at $\kappa = -3.25$.
We note that due to our periodic boundary conditions, the GBs always close on themselves.
There is, however, a qualitative difference between a situation in which a domain of one orientation is fully enclosed by another, as in Fig.~\ref{GrainBoundaries}(b), and one in which the domains wrap around the entire system.
In the latter case, the number of GBs is always even.

Among the possible GBs between triangular lattices\cite{Hirth82} we observe only the simplest kind, decorated with edge dislocations similar to those shown in Fig.~\ref{EdgeDislocations}.
The energy heat maps in Fig.~\ref{GrainBoundaries} indicate that just as in the case of individual dislocations, fivefold-coordinated vortices have higher than average energy and sevenfold-coordinated vortices have lower than average energy.
The lower dislocation density in Fig.~\ref{GrainBoundaries}(c) is in qualitative agreement with Frank's formula\cite{Frank50}, which predicts that the density of dislocations decorating the GB is proportional to the split angle $\Delta\theta$.
Here we have identified the GB between individual edge dislocations based on the change in the angle of the vortex lattice planes, as indicated by the white line in Fig.~\ref{GrainBoundaries}(c).

\begin{figure}
  \includegraphics[width = \columnwidth]{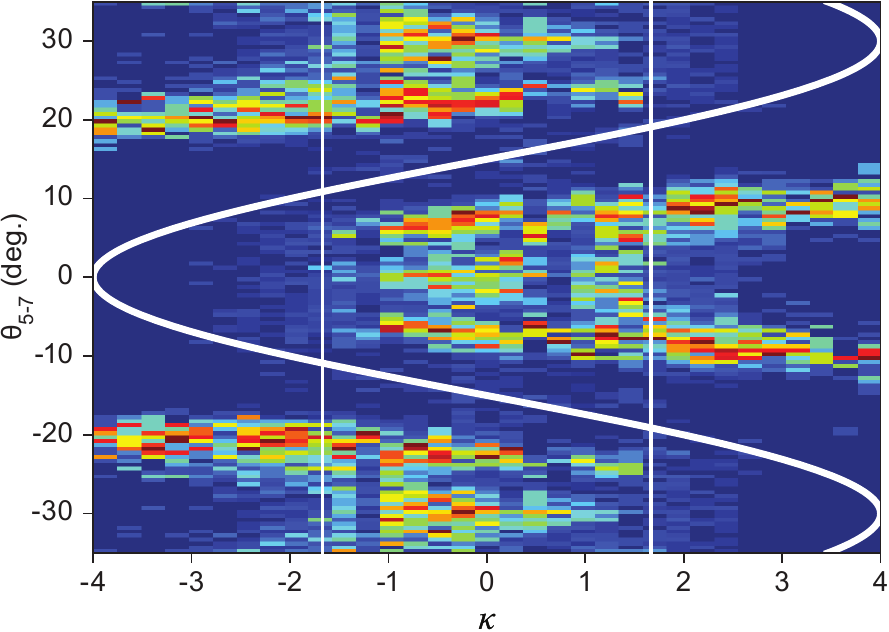}
    \caption{\label{GBhistogram}
      Heatmap showing the frequency of $\theta_{5-7}$, the angle of the line connecting the five-fold and seven-fold coordinated vortices in grain boundary edge dislocations, as a function of $\kappa$.
      All values of $\theta_{5-7}$ are mapped into a single $60^{\circ}$ segment.
      The heavy white curve indicates $\theta_{\text{min}}$, and the vertical lines show the values of $\kappa$ corresponding to $\Sigma = 7$.}
\end{figure}

Although we find curved GBs in our simulations for all values of the anisotropy ratio within the range $[-4,4]$, straight GBs, such as the one illustrated in Fig.~\ref{GrainBoundaries}(a), appear only when $0.75 \leq |\kappa| \leq 2.25$.
From Coincident Site Lattice (CSL) theory, straight GBs are expected to be favored for specific values of $\Delta \theta$ that are commensurate with the lattice symmetry\cite{Sutton95}.
Here, the degree of fit $\Sigma$ between adjacent domains is defined as the ratio of the total number of sites to the coincidence sites, and GBs with low integer values of $\Sigma$ have the lowest energy and are expected to be the most stable.
For the two-dimensional triangular lattice relevant to this work, the two lowest values are $\Sigma = 7$, corresponding to $\theta_{\rm{min}} = 10.9^{\circ}$  or $49.1^{\circ}$ and $|\kappa| = 1.67$, and $\Sigma = 13$, corresponding to $\theta_{\rm{min}} = 13.9^{\circ}$  or $46.1^{\circ}$ and $|\kappa| = 0.46$.
Considering the rotation transition in Fig.~\ref{SystemSize}, there are no obvious features at these specific values of $\theta_{\rm{min}}$.
As we discuss further below, however, a cross-over that coincides with $\Sigma = 7$ appears for the orientation of the GB edge dislocations.
We analyze the GB by measuring the angle $\theta_{5-7}$ between the five- and seven-fold coordinated vortices with respect to the positive $x$-axis.
Using the six-fold symmetry of our system, we project $\theta_{5-7}$ into the range $[-30^{\circ},30^{\circ}]$.
In Fig.~\ref{GBhistogram} we plot the observation frequencies of $\theta_{5-7}$ for different values of $\kappa$, and find that the observations always fall between the minima in the interaction potential indicated by the heavy white line.
This is consistent with what we find for free edge dislocations in Fig.~\ref{EdgeDislocations}.
In addition, the data in Fig.~\ref{GBhistogram} falls into two distinct regimes:
for $1.67 < |\kappa| < 4$, $\theta_{5-7}$ is clustered at two distinct values, while for $0 < |\kappa| > 1.67$, we observe six peaks in the angle histogram.
The boundary between the two regimes coincides with $\Sigma = 7$.

\section{Discussion}
In this work we considered only the static properties of the VL.
A logical next step would be to examine the vortex dynamics.
For example, if the vortex-vortex interactions were suddenly changed in real time, such as by a rapid increase or decrease in the vortex density driven by a change in the applied magnetic field, then the VL structures would need to change dynamically.
In this case, the vortex motion would likely be dominated by the defect structures and grain boundaries or by plastic rearrangements.
This is directly related to the metastable VL phases observed in MgB$_2$\cite{Das:2012cf,Rastovski:2013ff}.
Here, the activation barrier associated with the transition to the equilibrium phases increases as the metastable volume fraction is reduced\cite{Louden:2019bq,Louden:2019wx,Louden:2019jn}, suggesting a work hardening of the VL due to the proliferation of GBs.
Another area for study is both the statics and dynamics across the transition between the six-fold and 12-fold regimes
in the presence of quenched disorder.
If disorder is present, it might be strong enough to induce the formation of an intermediate glassy phase near the transition.
Under an applied drive, there could be a strong effect of the transition on the depinning threshold, which might exhibit some type of peak effect phenomenon.
In isotropic vortex systems in the presence of strong disorder, the pinned state is often highly disordered or glassy; however, under an applied drive, a plastic flow phase can appear followed at higher drives by a dynamically induced transition into a moving lattice or moving smectic state \cite{Olson98a,Koshelev94,Reichhardt17}.
In the anisotropic system, it would be interesting to see whether the presence of strong pinning combined with a drive can produce similar dynamical ordering transitions, and whether the dynamically ordered state would be a lattice with six- or 12-fold ordering, a smectic state, or some completely different type of moving phase.

Our results can be generalized to other particle based systems.
For example, the expression in Eq.~(2) could be modified easily by replacing the Bessel function $K_0$ with a screened Coulomb interaction $e^{-\kappa r}/r$ of the type that describes charge-stabilized colloids\cite{Libal:2007tv,Pertsinidis:2001hz} or
with a $1/r^3$ interaction that describes  magnetically interacting colloids\cite{Tierno12a}.
One could also examine whether the melting transition would change.
For isotropic interactions in a 2D system, the melting transition can occur in two stages through the Kosterlitz-Thouless-Halperin-Nelson-Young (KTHNY) mechanism\cite{vonGrunberg04,Kosterlitz73,Young79,Nelson79,Strandburg88}, which includes an intermediate hexatic phase, or it can be a first order process mediated by grain boundaries\cite{Strandburg88,Chui83}.  
In the KTHNY case, the hexatic phase often occurs only over a very small range of parameters or temperatures\cite{Marcus96,Zahn99}. 
If the system has a strong six-fold anisotropy, the hexatic phase could be strongly enhanced.

Finally, we notice that the interaction potential in Eq.~(\ref{K612potential}) should ideally be replaced by the results of a microscopic {\em ab-initio} calculation, both in terms of the angular dependence and the radial-angular decomposition.
This is outside the scope of the current work, and is left as a motivation for future work.

\section{Summary}
We have expanded our previous MD simulations of vortices by incorporating a combined six-fold and 12-fold anisotropy in the pairwise interaction potential.
Using this model we are able to reproduce the continuous $30^{\circ}$ rotation transition of the triangular VL that has been observed experimentally in superconducting MgB$_2$ and UPt$_3$.
We observe a spontaneous formation of dislocations as well as vacancy and interstitial defects, and characterize these in terms of their structure and energy distribution.
Grain boundaries separating differently oriented VL domains and decorated by edge dislocations, appear as the rotation transition progresses.
Large values of the anisotropy produce cluster crystal states.
Our model could be applied to other particle-based systems, such as magnetic skyrmions or colloids with anisotropic interactions, by modifying the isotropic contribution to the pairwise interactions.

\section*{Acknowledgements}
We are grateful to P.~Carstens, D.~Green, M.~Lamichhane, E.~R.~Louden, N.~Leisen, X.~Ma, D.~McDermott and K.~Newman for assistance and discussions.
This research was supported in part by the Notre Dame Center for Research Computing.
Work at the University of Notre Dame (MWO, MRE: MD simulations, data analysis) was supported by the US Department of Energy, Office of Basic Energy Sciences, under Award No. DE-SC0005051.
Part of this work (CR, CJOR: code development) was carried out under support by the US Department of Energy through
the Los Alamos National Laboratory.  Los Alamos National Laboratory is operated by Triad National Security, LLC, for the National Nuclear Security Administration of the U. S. Department of Energy (Contract No. 892333218NCA000001).

\section*{References}
%

\end{document}